# Simultaneous measurements on the electron and X-ray spectra from laser-irradiated near-critical-density double-layer targets at relativistic intensity

Jianbo Liu[1], Pengjie Wang[1], Yinren Shou[1], Zhusong Mei[1], Zhengxuan Cao[1], Zhuo Pan[1], Defeng Kong[1], Shirui Xu[1], Guijun Qi[1], Zhipeng Liu[1], Shiyou Chen[1], Jiarui Zhao[1], Yanying Zhao[1], Wenjun Ma[1,2,3]*

[1]School of Physics, State Key Laboratory of Nuclear Physics and Technology, Peking University, Beijing 100871, China

[2]Beijing Laser Acceleration Innovation Center, Huairou, Beijing,101400, China

[3]Institute of Guangdong Laser Plasma Techology, Baiyun, Guangzhou,510540, China

* Email: wenjun.ma@pku.edu.cn

Abstract:

We report the experimental results of simultaneous measurements on the electron and X-ray spectra from near-critical-density (NCD) double-layer targets irradiated by relativistic femtosecond pulses at the intensity of $5 \times 10^{19}\ W/cm^2$. The dependence of the electron and X-ray spectra on the density and thickness of the NCD layer was studied. For the optimal targets, electrons with temperature of 5.5 MeV and X-rays with critical energy of 5 keV were obtained. 2D particle-in-cell simulations based on the experimental parameters confirm the electrons are accelerated in the plasma channel through direct laser acceleration, resulting in temperature significantly higher than the pondermotive temperature. Bright X-rays are generated from betatron emission and Thomson backscattering before the electrons leave the double-layer targets.

Main Text:

With the advent and development of high-power laser technology, laser-driven electron acceleration in plasma has made great progress [1]. GeV level collimated electron beams have been generated from underdense plasma [2-4]. Oscillated by the electromagnetic field in plasma or by external magnets, such energetic electrons can produce brilliant X-rays with ultrashort durations and micrometer-scale source sizes, which are highly demanded in X-ray phase-contrast imaging [5] and X-ray radiography of high-energy-density matters [6]. Recently, all-optical controllable laser-plasma X-ray/gamma-ray sources have attracted much attention. KeV to MeV energy-tunable X-ray and gamma-rays can be generated through Thomson backscattering [7-10] or betatron emission [11-14] utilizing laser-wakefield accelerated electrons. However, the number of electrons accelerated in the bubble regime is difficult to exceed tens of pC per shot, which typically results in a photon number of $10^7$/shot and energy conversion efficiency of $10^{-5}$. For many applications requiring high photon number, other electron acceleration and radiation generation schemes need to be explored.

One of the promising ways to increase the yield of the electrons and the X-rays is to achieve the resonance between the betatron oscillation of the electron in a plasma channel and the laser field, which happens most efficiently in NCD plasma [15-17]. Numbers of theoretical works have predicted that nC, 100s MeV electrons, and ultra-high brightness X-ray/gamma-ray radiations can be generated from the laser NCD plasma interactions by virtue of betatron emission [18-20] or Thomson backscattering [9, 21-25].

However, only limited experimental results were reported up to now due to the difficulties on the fabrication of NCD targets. One of the earliest results using sub-critical high-pressure gas jet demonstrated that directly laser accelerated (DLA) electrons have the characteristics of large divergence angle and Boltzmann-like energy spectrum up to 12 MeV. The energy conversion efficiency from laser to electrons can reach 5%[26]. Kneip et al. measured the electron and x-ray energy spectra from sub-critical gas jet irradiated by 100J/630fs laser pulses. They found the optimal electron density of the gas jet for high-energy electron generation is about $10^{19}$ cm$^{-3}$~$0.01n_c$, where $n_c = m_e\omega_0^2/4\pi e^2$ is the critical density. They concluded the measured X-rays up to 50 keV were well described in the synchrotron asymptotic limit of electrons oscillating in a plasma channel [27]. In 2013, Chen et al. utilized 3TW laser pulses interacting with Ar cluster targets and produced electron beams with charge up to 200pC and energy up to 30MeV. The measured brightness of the betatron X-rays was $10^{21}$ $photons/s/mm^2/mrad^2/0.1\%BW$ [28]. Rosmej et al. demonstrated that electron beam with charge up to 100nC and laser-to-electron energy conversion efficiency of 30% can be achieved when 100J/750 fs laser pulses interact with pre-ionized foam targets at intensity of 2-5× $10^{19}$ $W/cm^2$. High-dose bremsstrahlung radiation was generated when the electrons passed through a high-Z mental target[29]. These experiments proved that relativistic laser pulse interaction with NCD plasma in DLA regime could generate high-energy, high-charge electron beams as well as intense X-rays. However, systematic experimental studies with well-controlled target parameters, especially for $n_e \sim n_c$, are stilling lacking. Recently, carbon nanotube foams (CNF) were successfully employed as targets for enhanced ion accelerations[30, 31]. Ionized and heated by the rising edge of relativistic laser pulses, as-prepared CNFs will evolve into homogeneous plasmas with electron density from $0.2n_c$ to $2n_c$, and generate super-ponderomotive electrons. Since the density and thickness of CNFs can be well-controlled in the fabrication process, systematic studies and optimization of the electrons acceleration and X-ray generations in NCD plasma becomes convenient.

In this paper, we present the experimental results of simultaneous measurements on the electron and X-ray energy spectra from CNT-coated double-layer targets irradiated by relativistic femtosecond pulses at the intensity of $5 \times 10^{19} W/cm^2$. By varying the densities and thicknesses of CNF targets, the dependence of the electron temperature and X-ray photon energy on the target parameters was studied. 2D particle-in-cell simulations illustrate the electron acceleration process in the targets. The simulated electron temperatures after the sheath fields fit the experimental results very well. The electrons are most energetic before they leave the double-layer targets, producing bright X-rays from their betatron motions in the channel and Thomson backscattering in front of the second-layer target.

The experiments were performed at the laboratory of Compact Laser Plasma Accelerator (CLAPA) at Peking University. The CLAPA laser is a Ti: sapphire laser system with central wavelength $\lambda =$ 800nm and full-width at half-maximum (FWHM) duration of 30fs. S-polarized laser pulse with the energy of 1.8 J was first focused on a single plasma mirror to improve the contrast to $10^8$ @5ps, then recollimated again and sent to the target chamber with an energy of 1.2 J. Figure 1 shows the experimental setup in the target chamber. The laser beam was focused with an f/3 off-axis parabolic (OAP) mirror and normally shot the targets. The intensity distribution in the focal plane was measured by a microscopic imaging system equipped with a 12-bit charged coupled device (CCD). In the focal plane, the FWHM diameter of the focal spot was $w_0 \sim 5\mu m$, and 32% of the laser energy was concentrated in the FWHM area. Calculated from the pulse duration and the focal spot measurement, the peak laser intensity $I_0$ is $\sim 5 \times 10^{19}\ W/cm^2$, corresponding to normalized vector potential amplitude of $a_0 = eE/m_e\omega_L c \sim 5$. In this experiment, we used CNFs as NCD targets. Two kinds of CNFs with electron density of $0.7n_c$ and $0.2n_c$ were used in the experiment. Their thicknesses varied from 20 μm to 80 μm. 0.5μm Fomval foils were used as supporting foils for the CNFs. The energy spectra of the electrons were measured by a magnetic dipole spectrometer with an average field of 0.29 tesla. A 2cm-thick Teflon with a 3mm hole was positioned 30 cm away from the targets as the collimator. A front-covered Biomax phosphor screen was positioned to one side of the magnet to measure the electron distribution after they were deflected by the magnet. The signals on the phosphor screen were used to deconvolute the energy spectra. The X-rays passed through the collimator were measured by MS image plates (IPs) outside the chamber after a 300 μm Be window. A filter cake made of different metal foils (20 μm Al, 20 μm Cu, 40 μm Cu) was put in front of the image plates to obtain the X-ray spectra. The X-ray response curves after the filters were obtained by multiplying the transmittance function of the filters by the IPs response function[32]. The distance from the targets to the image plates is 67 cm, resulting in a reception angle of 2.2×10⁻⁶ sr/mm². The drift distance of 24 cm from the magnet to the IPs guarantee that electrons below 80 MeV would not hit on the IPs.

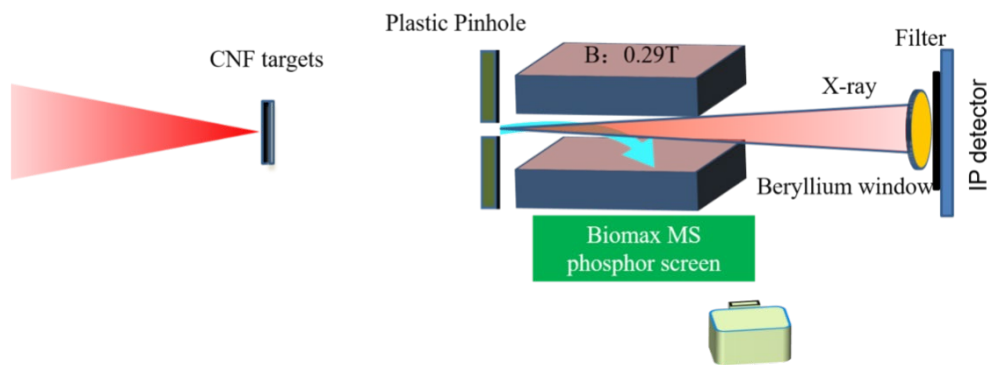

Figure 1. Schematic of experiment setup

Figure 2(a) shows the raw data and the derived electron spectrum for a 60 μm $0.2n_c$CNF+ 0.5 μm CH target. The electron spectrum has a Boltzmann-like distribution extending to 10 MeV with a fitted temperature of 3.4 MeV. Figure 2(b) shows the photostimulable luminescence (PSL) data of the X-rays measured by the IPs behind the filter cake. Filter No.1-4 represents 300 μm Be, 300 μm Be+20 μm Al, 300 μm Be+20 μm Cu, 300 μm Be +40 μm Cu, respectively. The transmissions of the filters in the range of 1 keV-30 keV are shown in figure 2(c). We define the cut-off energy of the filters by the energy where the transmission is 1%. The corresponding cut-off energy for filter No.1-4 is about 2.5 keV, 3.6 keV, 4.7 keV, 6.5 keV, respectively. One can see that the X-ray signals drop dramatically for the 4$^{th}$ filter, which indicates that the X-ray photons are mainly below 6.5 keV. We performed least-squares fits by adopting the synchrotron spectrum and thermal spectrum as the tentative functions. The fitted critical energy and temperature is 3.2 keV and 3.6 keV, respectively. Considering the two methods give similar values, we use the critical photon energy $E_c$ to represent the energy of the X-ray photons in this work. The fitted the spectra is displayed in figure 2(d). The x-ray yield is $1.8 \times 10^9 photons/Sr$ by integrating the whole spectra range.

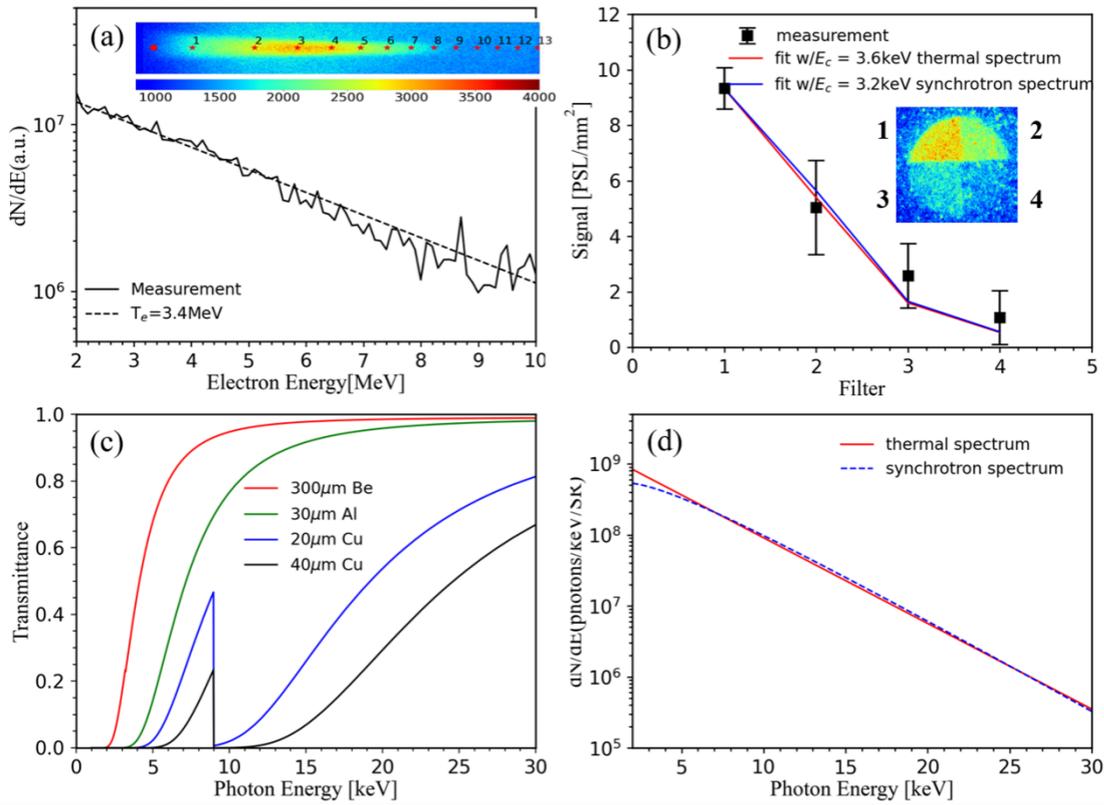

Figure 2. A typical experiment result obtained from a 60 μm CNF ($0.2n_c$) + 0.5 μm CH target. (a) Raw data (inset figure) and the energy spectrum of the electrons. (b) Raw data (inset figure) and the PSL value (black square dots) of the IPs behind the filters. The red triangle dots depict the fitted thermal spectrum with $E_c = 3.6\ keV$ and the blue triangle dots depict the fitted synchrotron spectrum with $E_c = 3.2\ keV$. (c) The transmittance of the filters. The propagation of the X-rays in 2 cm air is also considered. (d) The fitted energy spectra by adopting the thermal function or synchrotron function as the tentative function, respectively.

We did series experiments to study the dependence of the electron and X-ray spectra on the target parameters by varying the CNF's thickness (L) and electron density ($n_e$). The results are presented in Figure 3. Figure 3(a) and 3(c) shows the electron spectra data. It can be seen that all the electron spectra in the measurement range have the Boltzmann-like distribution, but the numbers and temperatures strongly rely on the target parameters. The $T_e$ of $0.2n_c$ targets are significantly higher than that of the $0.7n_c$ targets by a factor of 3-5, maximized at L=80 μm with $T_e = 5.5\ MeV$, and the electron numbers of $0.2n_c$ targets are also higher. For targets with $n_e = 0.2n_c$, $T_e$ and the electron numbers rise with increased L. Instead, for $0.7n_c$ targets, the dependence of $T_e$ on the thickness of the targets is weak, and the highest electron number was obtained for L=40 μm. Figure 3(b) and (d) show the X-ray data. It can be seen that, generally speaking, the photon numbers of the $0.7n_c$ targets are higher than that of the $0.2n_c$ targets. However, the critical photon energy $E_c$ of the $0.2n_c$ targets is higher instead. For targets with $n_e = 0.7n_c$, the photon number and $E_c$ is almost independent on the thickness of the targets. While for $0.2n_c$ targets, minimum $E_c$ was achieved for L=60 μm, and the highest $E_c$ of 5 keV was obtained for L=80 μm.

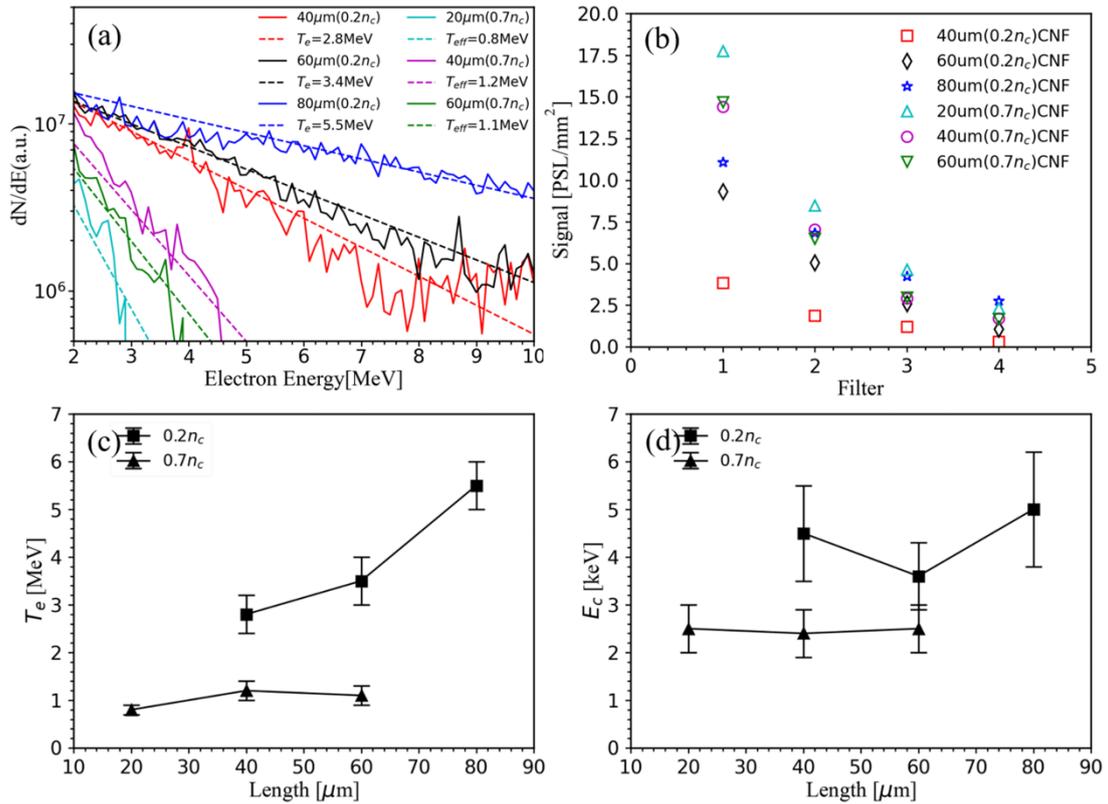

Figure 3. Experimental results on the electron and X-ray measurement. (a) Measured energy spectra of the electrons (solid lines) and the exponential fits (dotted lines). (b) Raw PSL data from the IPs behind filters. Electron temperatures (c) and critical photon energies (d) for targets with varied thicknesses.

By looking into the experimental results, we can make interpretations as below. First of all, the experimental results unambiguously show that $n_e = 0.2n_c$ targets are superior to the $n_e = 0.7n_c$ targets for the generation of high-energy electrons at $a_0 \sim 5$. In the near-critical plasma channel, the resonant direct laser acceleration happens when the electrons' betatron frequency equals to the laser frequency in the electron comoving frame. It was revealed in previous studies that [33, 34] the electrons gain the highest energy when $G = a_0\sqrt{n_e/n_c} \sim 1$. The G for $0.2n_c$ and $0.7n_c$ target is 2.2 and 4.2, respectively. The higher $T_e$ of $0.2n_c$ targets is in agreement with the theory. Secondly, for $0.7n_c$ targets, the $T_e$, $E_c$ and X-ray yield are similar for L=20$\mu m$, 40$\mu m$, 60$\mu m$, which indicates the electron acceleration and X-ray emission process are similar in those targets. Previous studies reveal that the dissipation length of laser in plasma $L_{etch} \approx \frac{3}{8}\frac{a_0}{n_e/n_c} c\tau$ [35]. Calculated from this formula, we can obtain $L_{etch} = 23$ $\mu m$ for $n_e = 0.7$ $n_c$. It means the laser pulses would dissipate their energy after propagating over 23 $\mu m$ in $0.7n_c$ CNF. This explains why the electron and X-ray spectra are similar for CNF with L=20-60$\mu m$. Thirdly, for $0.2n_c$ targets, their $T_e$ rises up with increasing L, however, the critical photon energy $E_c$ is minimal for L=60 $\mu m$. Such an inconsistency implies the dominant radiation schemes for L=20 $\mu m$ and L=60 $\mu m$ are different.

In order to confirm our interpretations and reveal the underlying physics, we performed a series of 2D PIC simulations with EPOCH code. The simulations utilized a fixed window of 75 μm in the laser propagation direction and 40 μm in the transverse direction. The longitudinal and transverse resolution is 60 cells/μm and 25 cells/μm, respectively. The driving laser's normalized amplitude $a_0$, duration τ, and laser wavelength $\lambda_L$, full-width-at-half-maximum of the focal spot of, was set as 5, 30 fs, 0.8 $\mu m$, 5 $\mu m$, respectively, to match the experimental parameters. A $200n_c$/0.5μm plasma is positioned at 60 μm to represent the CH foil. Uniform carbon plasmas with electron densities of $0.2n_c$ and $0.7n_c$ are positioned in front of the CH target to represent the CNF targets. The trajectories of test electrons are tracked.

Figure 4(a) and Figure 4(b) shows trajectories and longitudinal momentum $p_x$ of 10 test electrons for a 60 μm CNF (0.2 $n_c$) + 0.5 μm CH target. For most of the interactions, the electrons oscillate with small amplitudes of a few μm without picking up significant longitudinal momentum. After a while, some electrons gain large $p_x$ due to the resonant coupling with the laser field and meanwhile oscillate with amplitudes larger than 10 μm. When the electrons pass the second-layer plasma, they are decelerated in the sheath field and results in a drop of $p_x$. Figure 4(c) presents the evolution of $T_e$ with time for different targets. Here $T_e$ is the temperature of the forwardly electrons. For the comparison with the experimental data, only electrons with $p_\perp/p_x < 0.02$ and $p_x c > 2 MeV$ are counted [29]. $T_e$ for all the targets first rise up, then stay the same, eventually drops in the sheath field of the CH targets. We name the $T_e$ in the second and third stage as plateau temperature $T_{e\_pl}$ and final temperature $T_{e\_f}$, respectively. The $T_{e\_f}$ represents the electron temperature measured by the spectrometer and the plateau $T_{e\_pl}$ is more relevant to the X-ray emission. The $T_{e\_f}$ of the $0.2n_c$ targets range from 3.1 MeV to 5.5 MeV, which is in agree with the experimental results measured by our electron spectrometer. It should be noticed that the $T_{e\_f}$ of the 80 μm CNF ($0.2n_c$) + 0.5 μm CH target is the highest among all the targets. This is because the laser pulse in that case deposit most of its energy in the NCD plasma before it arrives at the CH target, resulting in a weaker sheath field. The $T_{e\_f}$ of the $0.7n_c$ targets are about 1 MeV, which again is in agree with the experimental results. The $T_{e\_pl}$ of $0.2n_c$ targets is about 7 MeV, which is 2 times of that of the $0.7n_c$ targets, and 7 times of the ponderomotive temperature. Such high-energy electrons are crucial for the generation of the X-ray radiations.

In the double-layer targets, the X-rays are generated in two mechanisms. The first is the betatron radiation resulted from the oscillation of the electrons in the channel. Approximately, the critical photon energy of the betatron radiation is[28]

$$E_{c\_B}[keV] = 5.3 \times 10^{-24} \langle \gamma_x^2 \rangle n_e[cm^{-3}] r_\beta[\mu m] \quad (1)$$

where $r_\beta$ is the amplitude of the oscillation motion. Considering the electron energy spectrum of $dN/dE \propto \exp(-E/T_e) \approx \exp(-\gamma_x/T_e)$, we have

$$\langle \gamma_x^2 \rangle = \int_0^\infty \gamma_x^2 \exp(-E/T_{e\_pl}) d\gamma_x / \int_0^\infty \exp(-E/T_{e\_pl}) d\gamma_x \approx 2 T_{e\_pl}^2 \quad (2)$$

where $T_{e\_pl}$ could be obtained in figure 4 (c). $r_\beta$ approximately equals to 5μm as shown in Figure 4(a). Then we can make a theoretical prediction of $E_{c\_B}$ for different targets. Figure 4(d) shows the comparison between the experimental results and the predicted $E_{c\_B}$ according to Eq (1), It can be seen that the experimental and predicted curves fit each other very well for $0.7n_c$ targets. However, they do not completely match for $0.2n_c$ targets. The experimental $E_c$ for the L=40 μm target is significantly lower than that of the predicted $E_{c\_B}$. We believe this is because the second radiation mechanism, Thomson backscattering, dominates for such a target. The dissipation length $L_{etch}$ of $0.2n_c$ target is 84μm. For the 40 μm CNF (0.2 $n_c$) + 0.5 μm CH target, the laser pulse still contains 50% of its original energy when it arrives at the CH foil. The reflected pulse will collide with the DLA electrons in the channel and generate X-rays in the forward direction[36]. The photon energy of the X-rays $E_X = 4a_{rf}\gamma_x^2 E_{ph}$ if $a_{rf} \gg 1$ or $E_X = 4\gamma_x^2 E_{ph}$ if $a_{rf} < 1$, where $a_{rf}$ is the normalized amplitude of the reflected pulse. We adopted the values of $a_{rf}$ from simulations and plotted the $E_X$ in figure 4(d). One can see it well explains the experimental $E_c = 4.5\ keV$ for L=20 μm. It should be noted that the Thomson backscattering has more potential in terms of generating high-energy photons as compared to betatron radiation. With the increase of $a_0$, $\gamma_x$ can be as high as 100s and the $a_{rf} \gg 1$, which would boost the photon energy to gamma-ray range.

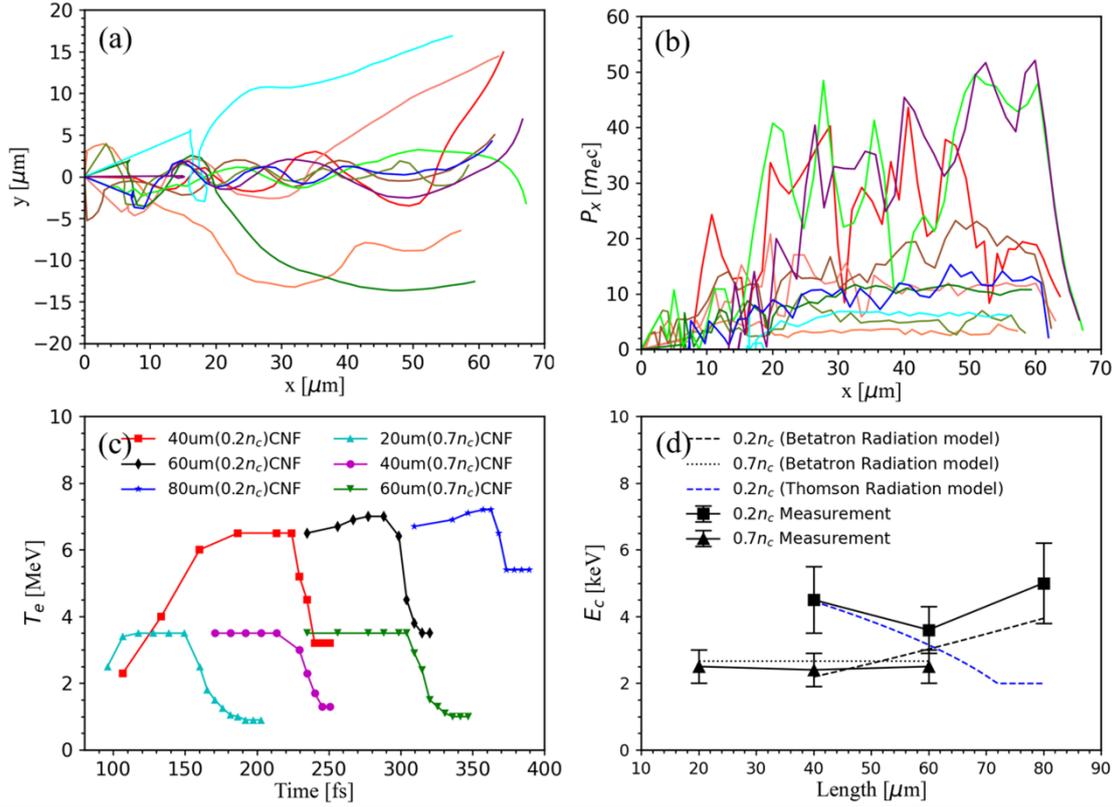

Figure 4. Results of 2D PIC simulations. (a) Trajectories of the 10 tracked electrons obtained from the 2D simulations, and (b) longitudinal momentum $p_x$ for the target of 60 μm CNF ($0.2n_c$) + 0.5 μm CH. (c) Evolution of $T_e$ with time for different targets. (d) Comparison of the experimental and theoretical values of $E_c$ for X-ray photons.

In summary, we simultaneously measured the electron and the X-ray energy spectra from laser-irradiated $0.2n_c$ and $0.7n_c$ near-critical-density double-layer target at intensity of $5 \times 10^{19}\ W/cm^2$. Electrons with temperature of 1-5.5 MeV, and X-rays with critical energy of 2.5 keV-5 keV were obtained depending on the target parameters. Series 2D PIC simulations based on the experimental parameters confirm that electrons in our targets are accelerated in NCD plasma channel through direct laser acceleration mechanism, resulting in temperatures significantly higher than the ponderomotive temperature. The betatron motion of the electrons leads to the X-rays emission in all the targets. Additionally, Thomson backscattering happens when the NCD targets are thinner than the dissipation length of the laser pulses. Our results provide a fresh and complete dataset for the study of the electron acceleration and X-ray generation in NCD plasma at relativistic intensity, which is also a valuable reference for studies aiming for the generation of bright gamma rays at higher intensity.

Acknowledgement: The work was supported by NSFC innovation group project (11921006), National Grand Instrument Project(2019YFF01014402), Natural Science Foundation of China (Grant No. 11775010,11535001, 61631001). The simulations are supported by High-performance Computing Platform of Peking University.